# Heterogeneous electrocatalysis in porous cathodes of solid oxide fuel cells

Y. Fu[1], S. Poizeau[2], A. Bertei[1,3], F. C. Qi[2], A. Mohanram[2], J. D. Pietras[2], M. Z. Bazant[1,4]

[1] Department of Chemical Engineering, Massachusetts Institute of Technology, Cambridge, MA

[2] Saint-Gobain Ceramics and Plastics, Northboro Research and Development Center, Northboro, MA

[3] Department of Civil and Industrial Engineering, University of Pisa, Italy

[4] Department of Mathematics, Massachusetts Institute of Technology, Cambridge, MA

## Abstract

A general physics-based model is developed for heterogeneous electrocatalysis in porous electrodes and used to predict and interpret the impedance of solid oxide fuel cells. This model describes the coupled processes of oxygen gas dissociative adsorption and surface diffusion of the oxygen intermediate to the triple phase boundary, where charge transfer occurs. The model accurately captures the Gerischer-like frequency dependence and the oxygen partial pressure dependence of the impedance of symmetric cathode cells. Digital image analysis of the microstructure of the cathode functional layer in four different cells directly confirms the predicted connection between geometrical properties and the impedance response. As in classical catalysis, the electrocatalytic activity is controlled by an effective Thiele modulus, which is the ratio of the surface diffusion length (mean distance from an adsorption site to the triple phase boundary) to the surface boundary layer length (square root of surface diffusivity divided by the adsorption rate constant). The Thiele modulus must be larger than one in order to maintain high surface coverage of reaction intermediates, but care must be taken in order to guarantee a sufficient triple phase boundary density. The model also predicts the Sabatier volcano plot with the maximum catalytic activity corresponding to the proper equilibrium surface fraction of adsorbed oxygen adatoms. These results provide basic principles and simple analytical tools to optimize porous microstructures for efficient electrocatalysis.

## 1) Introduction

Many important electrochemical reactions require electrocatalysts that accelerate the kinetics, while remaining unaltered. For Faradaic reactions at electrodes, electrocatalysis is heterogeneous since charge transfer occurs at the interface between electrode and electrolyte phases. Heterogeneous reactions typically involve multistep processes of consecutive and parallel elementary reaction steps, such as adsorption of chemical species on the electrode surface, surface or bulk diffusion of intermediate species, and charge transfer reactions at the electrode-electrolyte interface. An electrocatalyst accelerates the global reaction rate by lowering the barrier for the slowest elementary step in the reaction mechanism, and it is important to recognize that this might not necessarily be a charge transfer step.

The oxygen reduction reaction (ORR) in different types of fuel cell cathodes typically requires an electrocatalyst. For room-temperature proton-exchange-membrane fuel cells (PEMFC), platinum-based electrocatalysts are used and have been described by models that emphasize the charge transfer step [1,2]. Here, we focus on high-temperature solid oxide fuel cells (SOFC) with conducting ceramic electrocatalysts and also consider the intermediate steps of dissociative surface adsorption and surface diffusion. The overall reaction, written in Kröger–Vink notation, is as follows [3]:

$$\frac{1}{2}O_{2(g)} + 2e^- + V_O^{\bullet\bullet} \rightarrow O_O^x \quad (1)$$

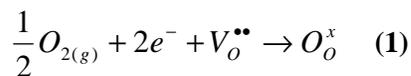

Eq. (1) highlights the need for a catalytic electrode able to act simultaneously as a good electronic conductor, ionic conductor and gas distributor in order to minimize the cathode overpotential, which is one of the major resistances in hydrogen-fed SOFCs [3]. These requirements are usually met by fabricating a porous composite electrode through the mixing of particles of a good catalytic electronic conductor, such as strontium-doped lanthanum manganite (LSM), and a good ionic conductor, such as yttria-stabilized zirconia (YSZ).

Since several species take part to Eq. (1), the ORR is likely to be the result of a multi-step reaction mechanism. For the LSM/YSZ electrode, two parallel reaction pathways have been proposed [4], namely the 'surface path' and the 'bulk path'. In the both the mechanisms, the first step consists in the adsorption and dissociation of oxygen molecules onto the LSM surface. Then, in the surface pathway, adsorbed intermediates diffuse on the LSM surface and reach the triple phase boundary (TPB) among LSM, YSZ and gas phases, wherein they undergo a charge transfer reaction with incorporation of oxygen ions into the electrolyte. On the other hand, the bulk pathway assumes that adsorbed intermediates are first incorporated within the LSM lattice, then transported within the bulk to the LSM/YSZ interface wherein the oxygen ion charge transfer completes the reaction mechanism.

Several kinetic studies have been performed in microstructured thin film electrodes in order to elucidate the relative importance of surface and bulk paths and to identify the corresponding rate-determining steps [5–10]. However, at the present there is no general agreement: some studies found that the kinetics is proportional to the TPB length per electrode area [11–13], thus suggesting a surface reaction mechanism, while others revealed a proportionality with the contact area between LSM and YSZ [7,14,15], which is an indication supporting the bulk pathway. Furthermore, several works indicate that both the surface and the bulk paths contribute together to the overall current [7,8,14,16]. For example, Horita et al. [6], who used isotope oxygen exchange and secondary ion mass spectrometry analysis, showed that the whole LSM/YSZ interface is active for high cathodic overpotentials, while only a finite extension around the TPB takes part to the ORR under low polarization. Similarly, la 'O et al. [10] concluded that the overall ORR rate may be limited by mixed bulk/surface charge transfer processes below 700°C and by surface chemical reactions above 700°C.

These experimental results have been often interpreted as the ORR was limited by a charge transfer step located, respectively, at the TPB or at the LSM/YSZ contact area. However, this is not necessarily the case. Adler and other researchers pointed out that, even though the phenomenological response suggests a charge transfer limitation (for example, a Butler-Volmer-like kinetics), such a behavior can be mimicked by slow transport and chemical steps without involving any charge transfer [3,17,18]. For example, even when interfacial charge transfer steps are equilibrated, the diffusion of a reaction intermediate can give rise to a Tafel behavior due to the logarithmic Nernstian relationship between the applied potential and the intermediate activity [3].

This conclusion is supported by several experimental studies. Van Heuveln et al. [19,20] reported that the surface diffusion of oxygen species on the LSM competes with charge transfer at the TPB, and the former is the rate-determining step at low overpotential. Østergård and Mogensen [21] revealed three features in the impedance spectra of LSM/YSZ cathodes, which were associated to dissociative adsorption of molecular oxygen, diffusion of oxygen ions on the LSM/gas interface and transfer of oxygen ions into the electrolyte at the TPB, indicating that the first two processes may be the rate-determining steps under typical conditions. This result was confirmed by Siebert et al. [14], who suggested that the rate-determining step of the ORR at low overpotential is dissociative adsorption, and by Wang et al. [22], who indicated that intermediate frequency arcs in cathode impedance response are related to the dissociative adsorption of $O_2$.

Therefore, although different numbers of rate-determining processes were identified [23–28] and there is still discrepancy in their relative contribution and activation energies [14,29–34], the ORR seems to be limited by surface chemical phenomena rather than by interfacial charge transfer steps. In such a case, it is worth taking into account also the non-charge-transfer electrocatalytic steps, such as oxygen adsorption and diffusion, when modeling the ORR. Nowadays this problem can be addressed through physics-based models [35] rather than through phenomenological equivalent circuits. In particular, mechanistic models based on physical conservation laws can be solved to simulate impedance spectra [36–38]. The comparison between simulated and experimental spectra allows the identification of reaction mechanisms and the inference of specific parameters of the electrocatalytic steps.

In this study, the electrocatalytic surface steps of the ORR, such as oxygen adsorption and surface diffusion, are modeled in composite LSM/YSZ cathodes for SOFCs. An analytical expression of the cathode impedance is obtained and compared with experimental spectra of symmetric cells with different porous electrode microstructures under a range of operating conditions. The model is able to successfully connect the electrocatalytic contribution to the overall impedance with geometrical features of the microstructure obtained by direct imaging and statistical analysis.

The paper is organized as follows. In Section 2 the analytical expressions of the electrocatalytic model are derived. In Section 3 the model is validated against experimental impedance data for different oxygen partial pressures and electrode microstructures. A thorough discussion of model results is reported in Section 4. Finally, the general conclusions of the study are summarized in Section 5.

## 2) Model development for the electrocatalysis in SOFC cathode

In this Section, a general mathematical model for heterogeneous electrocatalysis is developed and applied to describe the catalytic oxygen reduction reaction in the LSM/YSZ cathode functional layer of an SOFC. Due to the conditions analyzed in this study, i.e., low overpotential and temperature in the order of 800°C, the surface path is assumed to be the dominant reaction mechanism, in agreement with the literature [3,10,19,20,39]. Therefore, no contribution from the bulk path is considered.

Figure 1 illustrates the reaction mechanism of the ORR considered in this study through the schematic representation of a section of LSM surface in contact with a YSZ particle at a TPB. In the figure, $L_s$, which is a characteristic length of the microstructure called surface diffusion length, represents the typical semi-distance between two TPBs. Gaseous $O_2$ molecules dissociate and adsorb onto the LSM surface according to the adsorption reaction:

$$O_{2(g)} + 2\Diamond_{(s)} \to 2O_{(s)} \quad (2)$$

where $\Diamond_{(s)}$ represents a free adsorption site on the LSM surface and $O_{(s)}$ an oxygen adatom on the LSM surface. Oxygen adatoms diffuse on LSM surface to the TPB where the following charge transfer reaction occurs:

$$O_{(s)} + 2e^{-}_{(LSM)} + V^{\bullet\bullet}_{O(YSZ)} \to O^{x}_{O(YSZ)} + \Diamond_{(s)} \quad (3)$$

Note that the sum of reactions (2) and (3) gives the global ORR in Eq. (1).

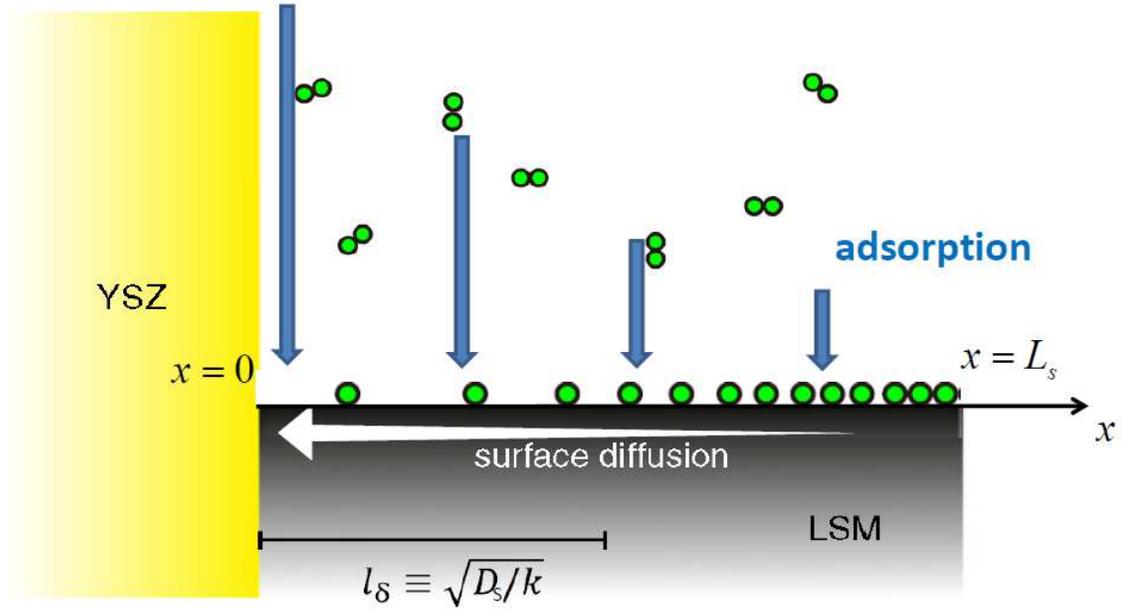

**Figure 1. Electrocatalytic kinetic process of LSM/YSZ porous electrode co-limited by surface diffusion and adsorption. Picture modified from Shin et al. [40].**

Assuming Langmuir dissociative adsorption kinetics for reaction (2) and Fick surface diffusion, the governing equation for oxygen adatoms on the LSM surface results as follows:

$$\frac{\partial \theta}{\partial t} = D_s \frac{\partial^2 \theta}{\partial x^2} + k_{ads}\left(K \cdot P_{O_2}(1-\theta)^2 - \theta^2\right) \quad (4)$$

where $\theta$ represents the dimensionless surface coverage fraction of oxygen adatoms, $D_s$ the surface diffusivity, $P_{O2}$ the oxygen partial pressure, $K$ the thermodynamic equilibrium constant of oxygen adsorption, and $k_{ads}$ the kinetic constant of the adsorption reaction.

When a sinusoidal external signal with angular frequency $\omega$ is applied around the open-circuit equilibrium, in the limit of small perturbations the surface coverage can be locally linearized as:

$$\theta = \theta_{eq} + \Delta\theta, \text{ with } \Delta\theta = \delta\theta \cdot e^{i(\omega t + \varphi)} \text{ and } \delta\theta << \theta_{eq} \quad (5)$$

where $i$ is the imaginary unit. Here $\theta_{eq}$ is the equilibrium coverage, which is a constant value resulting from the adsorption equilibrium as follows:

$$\theta_{eq} = \frac{\sqrt{K \cdot P_{O_2}}}{1 + \sqrt{K \cdot P_{O_2}}} \quad (6)$$

The substitution of Eqs. (5) and (6) into Eq. (4) yields:

$$i\omega\Delta\theta = D_s \frac{d^2\Delta\theta}{dx^2} - k\Delta\theta \quad \text{with } k = 2k_{ads}\sqrt{K \cdot P_{O_2}} \quad (7)$$

where terms in the order of $\Delta\theta^2$ are neglected.

Eq. (7) represents an ordinary differential equation which can be solved analytically upon the imposition of two boundary conditions, as follows:

$$\Delta\theta\big|_{x=0} = \theta_{eq}(1 - \theta_{eq})\frac{2F}{RT}\Delta V \quad (8a)$$

$$\frac{d\Delta\theta}{dx}\bigg|_{x=L_s} = 0 \quad (9b)$$

The first boundary condition (Eq. (8a)) represents the relationship between voltage perturbation and surface coverage perturbation at the TPB. Such a relationship comes from the application of the Nernst equation for reaction (3), which is assumed to be much faster (that is, in pseudo-equilibrium) than the adsorption and diffusion processes. The derivation of Eq. (8a) from the Nernst equation is reported in the Appendix. The second boundary condition (Eq. (8b)) represents a no-flux condition at the symmetry plane $x = L_s$ according to Figure 1.

The solution of Eq. (7) according to the boundary conditions Eqs. (8a) and (8b) provides the following profile of surface coverage perturbation:

$$\Delta\theta = \theta_{eq}(1-\theta_{eq})\frac{2F}{RT}\Delta V \frac{\cosh\left(\sqrt{\frac{k+i\omega}{D_s}}(L_s - x)\right)}{\cosh\left(\sqrt{\frac{k+i\omega}{D_s}}L_s\right)} \quad (10)$$

Eq. (9) is used to evaluate the impedance of the electrode. The current perturbation $\Delta I$ is obtained by considering that, according to reaction (3), two electrons are exchanged per each oxygen adatom that reacts at the TPB. Therefore, the current perturbation is proportional to the flux of oxygen adatoms at the TPB (i.e., at $x = 0$) as follows:

$$\Delta I = -2Fl_{TPB}C_{max}D_s\left.\frac{d\Delta\theta}{dx}\right|_{x=0} \Rightarrow$$

$$\Delta I = \frac{(2F)^2}{RT}l_{TPB}C_{max}D_s\theta_{eq}(1-\theta_{eq})\sqrt{\frac{k+i\omega}{D_s}}\tanh\left(\sqrt{\frac{k+i\omega}{D_s}}L_s\right)\Delta V \quad (11)$$

where $l_{TPB}$ is the triple phase boundary density per unit of electrode area and $C_{max}$ is the surface density of adsorption sites per unit of catalyst surface (LSM in this case). Note that in Eq. (10) a uniform overpotential perturbation $\Delta V$ is assumed throughout the whole cathode functional layer. However, there may be a distribution of overpotential across the electrode thickness due to electronic and ionic ohmic drops [41,42], especially for conditions in which a large current is drawn. The derivation of the impedance for a non-uniform distribution of overpotential is out of the scope of the paper and will be subject of a forthcoming publication.

Finally, the cathode impedance $Z_G$ is obtained by dividing the voltage perturbation $\Delta V$ by the current perturbation $\Delta I$ (Eq. (10)). After substitution of $\theta_{eq}$ as in Eq. (6), the area-specific impedance results as follows:

$$Z_G = \frac{\Delta V}{\Delta I} = \frac{RT}{(2F)^2}\frac{1}{l_{TPB}C_{max}D_s}\left(2+\sqrt{K\cdot P_{O_2}}+\frac{1}{\sqrt{K\cdot P_{O_2}}}\right)\frac{\coth\left(\sqrt{\frac{k+i\omega}{D_s}}L_s\right)}{\sqrt{\frac{k+i\omega}{D_s}}} \quad (12)$$

Eq. (11) is the impedance of an electrocatalytic process co-limited by dissociative Langmuir gas adsorption and surface diffusion associated to a charge transfer reaction at pseudo-equilibrium. It is interesting to note that Eq. (11) represents a Gerischer-like impedance element [43]. However, unlike the general impedance expression of a confined reaction-diffusion system reported by Sluyters-Rehbach and Sluyters [44,45], a hyperbolic co-tangent function appears in Eq. (11) instead of a hyperbolic tangent as a consequence of the no-flux boundary condition (Eq. (8b)). This feature was also reported by other authors in the literature, although for different applications [46,47].

The polarization resistance $R_G$ of the cathode can be calculated by setting the angular frequency to zero:

$$R_G = \frac{RT}{(2F)^2} \frac{1}{l_{TPB} C_{max} D_s} \left( 2 + \sqrt{K \cdot P_{O_2}} + \frac{1}{\sqrt{K \cdot P_{O_2}}} \right) \frac{\coth\left(\sqrt{\frac{k}{D_s}} L_s\right)}{\sqrt{\frac{k}{D_s}}} \qquad (132)$$

Eq. (12) allows the identification of a characteristic length $l_\delta$ of the adsorption/diffusion process, called boundary layer length, defined as follows:

$$l_\delta = \sqrt{D_s / k} \qquad (143)$$

As depicted in Figure 1, the adsorption and surface diffusion processes mainly occur within this boundary layer length in the proximity of the TPB. In addition, the ratio between the surface diffusion length $L_s$ and the boundary layer length $l_\delta$ identifies a dimensionless number, known as Thiele modulus [48]:

$$\phi = \frac{L_s}{l_\delta} = \frac{L_s}{\sqrt{D_s / k}} \qquad (154)$$

The Thiele modulus characterizes the electrocatalytic activity of the electrode for different electrode microstructures as discussed in Section 4.

In the next Section, the application of Eqs. (11-14) is discussed in comparison with the experimental impedance spectra of composite LSM/YSZ cathodes for different oxygen partial pressures and microstructures.

## 3) Validation of the cathode electrocatalysis model

### 3.1) Experimental methods

Symmetric button cells were fabricated in order to isolate the cathode contribution. Each electrode consisted of a porous bulk LSM layer (1.2mm thick) and a porous composite LSM/YSZ functional layer (15-25μm thick). A YSZ electrolyte (10-20 μm thick) was placed in between the two symmetric electrodes. Four different cells were fabricated, differing in the microstructure of the LSM/YSZ functional layer only. In all the cases, the cell diameter was 1 inch. The cells were sealed by using LP-1071 glass from Applied Technologies. Pt mesh was used on each side as current collector. Pt lead wires were connected from the current collector layers to the data collecting equipment.

A 1470E Solartron Analytical from MTechnologies and mSTAT program were used to control the operating conditions and collect the data. The air flow rate was 300 std cm$^3$/min on each side. The following protocol was adopted to measure the impedance. The cell was brought to 900°C overnight, then tested at 900, 850, and 800°C. After each temperature change, the cell was kept in open-circuit conditions for 1hr in order to reach equilibrium, which was confirmed by measuring the OCV. AC impedance data was collected at OCV for four different oxygen partial pressures (0.21, 0.15, 0.10, and 0.05atm), which were obtained by diluting air with nitrogen. At the end of the test matrix, the cell was brought back to 900°C to re-measure the first AC impedance dataset and confirm the reproducibility of the measurements.

### 3.2) Frequency and pO$_2$ dependence

Figure 2 shows the physics-based equivalent circuit adopted to fit the experimental impedance data. The inductance of lead wires is taken into account through an inductor $L$ connected in series to the cell impedance components according to Shin et al. [40]. The resistor $R$ represents the total ohmic resistance of the cell, mainly due to the electrolyte and contact resistances. The cathode impedance is represented by the two elements within the brackets. $G$ corresponds to the impedance of the surface adsorption/diffusion processes, whose analytical expression is reported in Eq. (11). The $R_F$ and $C_F$ in parallel account for the cathode charge transfer reactions, which are expected to contribute in the high frequency region. Note that the cathodic processes $G$, $R_F$ and $C_F$ are taken into account twice due to the symmetric configuration of the cell. Based on our previously validated, accurate model of gas diffusion in the same system [49], we can neglect the oxygen gas diffusion impedance to simplify the fitting.

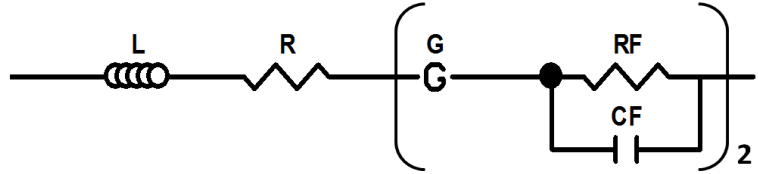

**Figure 2  Physics-based equivalent model for the symmetric cathode cells.**

Figure 3 shows the comparison between model results and experimental data for four different oxygen partial pressures. The cathode impedance shows a dominant depressed arc at low frequency, a Warburg-like behavior at high frequency and an inductive tail at very high frequency. Note that both the inductive tail and the ohmic resistance remain unaltered while varying the oxygen partial pressure, while the cathode polarization resistance decreases as $P_{O2}$ increases.

The model results accurately fit the experimental data across the whole frequency range for all the four oxygen partial pressures. The excellent agreement was obtained by fitting only eight unknown parameters: two of them, the ohmic resitance $R$ and the inductance $L$, were kept fixed for all the oxygen partial pressures, while six of them, whose values are summarized in Table 1, were left free for each $P_{O2}$. Note that, as reported in Table 1, the separate fitting of four curves give almost consistent values for $D_s$, $K$ and $k_{ads}$, which all should be material properties. The factor $l_{TPB}C_{\max}$, which depends on the microstructure of the porous electrode only, is almost constant as well, which further confirms the validity of the model.

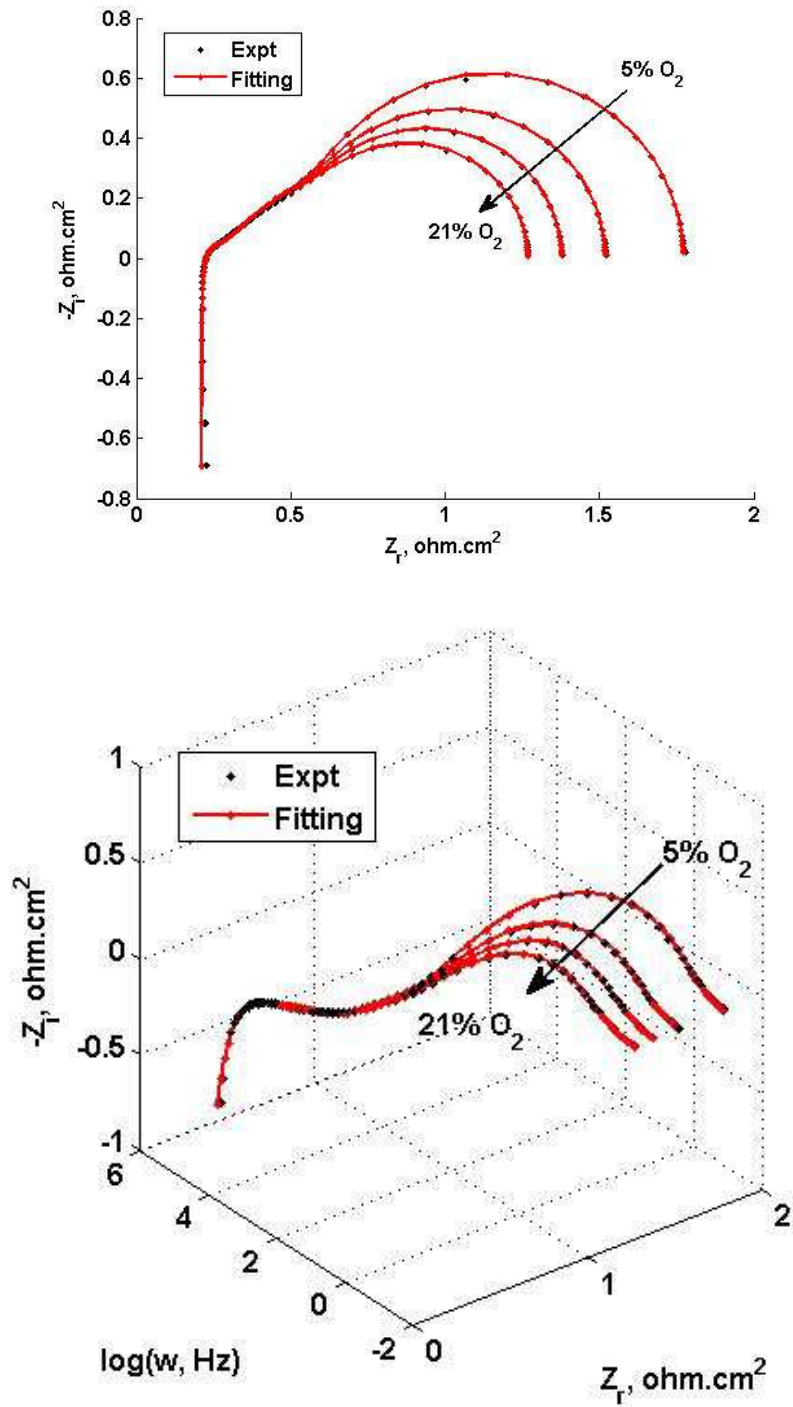

**Figure 3** Validation of frequency dependence of the proposed cathode model, when each curve at different oxygen partial pressure is fitted separately using the equivalent circuit reported in Figure 2. (Upper) 2D Nyquist plot. (Bottom) 3D plot showing frequency dependence.

**Table 1 Fitted parameters for Figure 3 where oxygen partial pressure in cathode feeding stream varies from 0.21 to 0.05atm.**

| p(O$_2$) | 0.21 | 0.15 | 0.10 | 0.05 | atm |
|---|---|---|---|---|---|
| $l_{TPB}C_{max}$ | 1.428E-2 | 1.261E-2 | 1.253E-2 | 1.293E-2 | mol/cm$^3$ |
| $D_s$ | 4.32E-9 | 4.26E-9 | 4.23E-9 | 4.18E-9 | cm$^2$/s |
| $K$ | 1.91E-4 | 6.05E-5 | 7.96E-5 | 4.73E-4 | 1/Pa |
| $k_{ads}$ | 1.722 | 3.179 | 2.953 | 1.389 | 1/s |
| $R_F$ | 0.0137 | 0.0138 | 0.0137 | 0.0137 | Ωcm$^2$ |
| $C_F$ | 0.0299 | 0.0274 | 0.0245 | 0.0212 | F/cm$^2$ |

Since material properties and microstructural parameters do not vary significantly with $P_{O2}$, their average values are fixed in order to further reduce the number of fitting parameters. Under this constraint, the fitting of the four experimental curves was performed again and reported in Figure 4. In this figure, only the high frequency resistance $R_F$ and capacitance $C_F$ were re-fitted for each $P_{O2}$. As shown in Figure 4, there is still a satisfactory agreement between experimental data and model results despite the reduced number of fitted parameters.

Hence, the results reported in this Section confirm that the proposed model is capable to capture both the frequency and oxygen partial pressure dependences of composite LSM/YSZ cathodes.

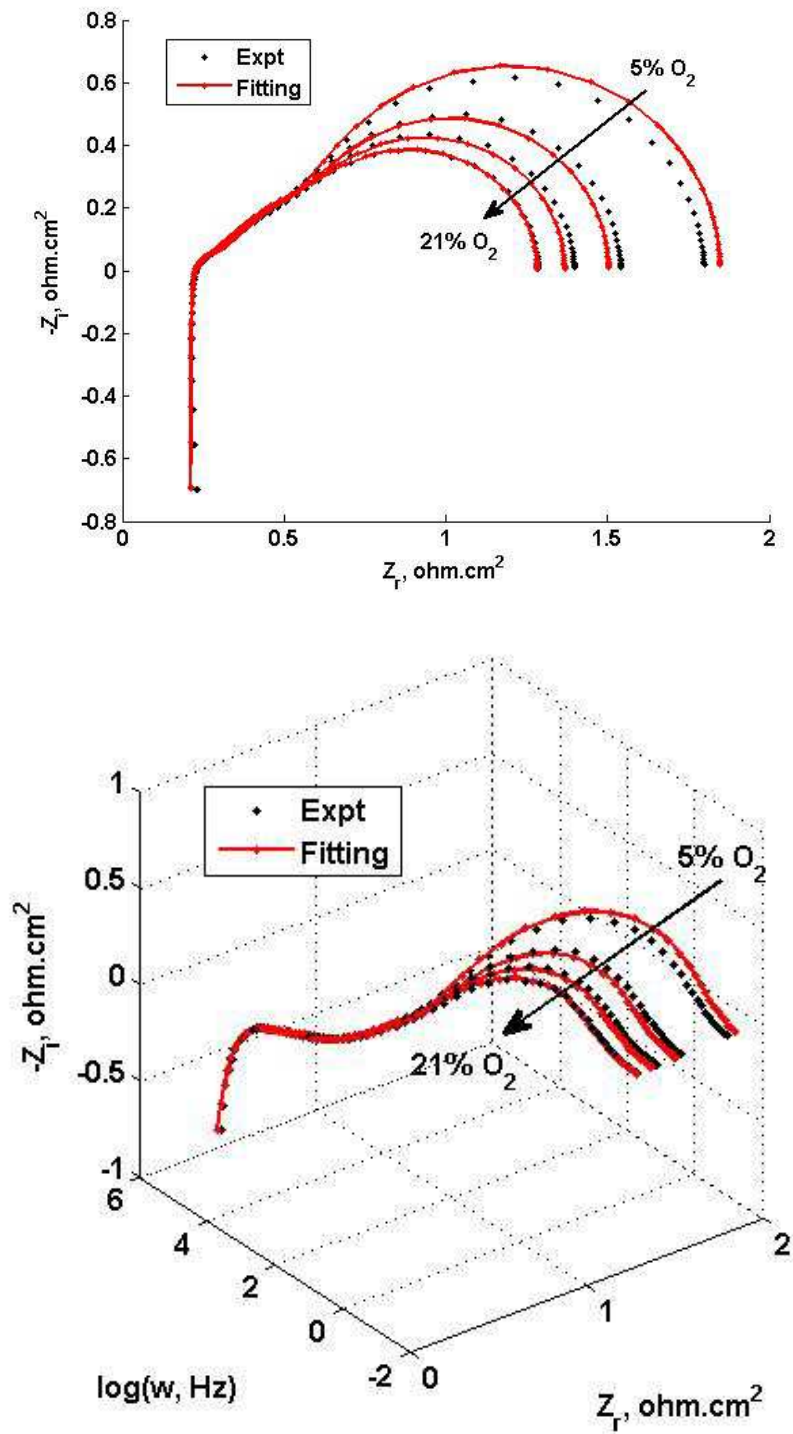

**Figure 4** Validation of oxygen partial pressure dependence of the proposed cathode model, when all the four curves are fitted together with fixed material property parameters in the equivalent circuit reported in Figure 2. (Upper) 2D Nyquist plot. (Bottom) 3D plot showing frequency dependence.

## 3.3) Microstructure of the cathode functional layer

In order to corroborate the validity of the model presented in Section 2, four different symmetric cells were fabricated (labeled as A, B, C and D in the following), differing in the microstructure of the cathode functional layer only. The materials and solid volume fractions of the cathode functional layer were the same for all the samples, while porosity and particle sizes were varied in order to obtain different final microstructures.

Each microstructure was characterized through the analysis of six SEM images per sample. An example of SEM image is reported in Figure 5. Each image was analyzed by using the image processing software ImageJ, which allowed the quantitative estimation of the number of TPB points per unit area $n_{TPB}$ and of the surface diffusion length $L_s$ as the median LSM/pore interface length divided by two. Note that the SEM images did not provide any information about the percolation properties of the gas and solid phases. This means that all the possible TPB points and surface paths, regardless of the real percolation status of pore, LSM and YSZ, were considered in the estimation of $n_{TPB}$ and $L_s$, respectively. Table 2 summarizes the microstructural properties of the four cathodes as estimated from the image analysis.

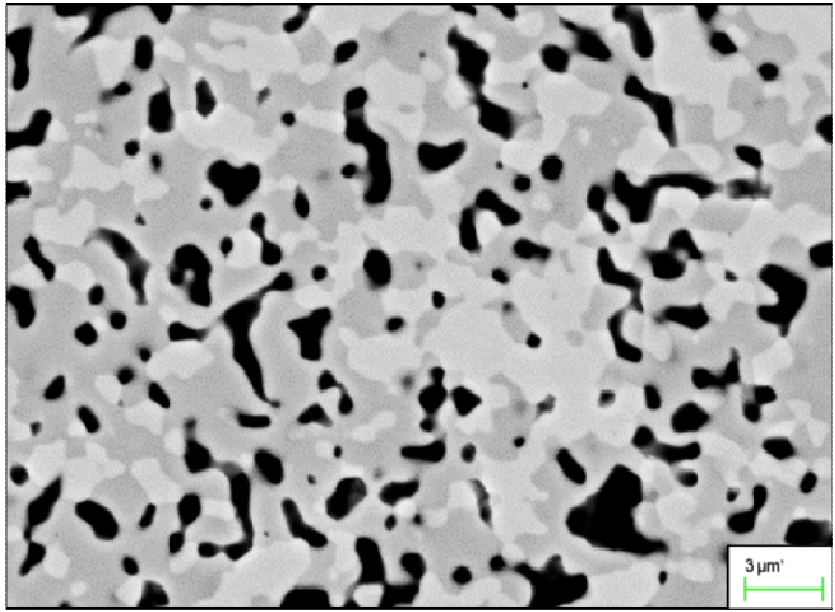

**Figure 5 A typical SEM image of the cathode functional layer. (Dark grey–YSZ, light gray–LSM, black–Pore)**

**Table 2 Microstructural parameters of the four symmetric cathode cells estimated through the SEM image analysis. Boundary layer length values, calculated from the fitting of impedance data at $P_{O2}$ = 0.21atm, are added in the last row.**

|  | Cell A | Cell B | Cell C | Cell D | Units |
|---|---|---|---|---|---|
| **Median $L_s$** | 0.29±0.03 | 0.41±0.03 | 0.31±0.07 | 0.20±0.06 | µm |
| **Median $n_{TPB}$** | 0.23±0.04 | 0.32±0.06 | 0.33±0.04 | 0.32±0.11 | 1/µm² |
| $l_\delta$ **at 21% $O_2$** | 0.30 | 0.29 | 0.22 | 0.36 | µm |

Impedance spectra at OCV and $P_{O2}$ = 0.21atm were recorded for each symmetric cell. Experimental data are reported in Figure 6. The figure shows that the curves exhibit the same depressed shape with Warburg-like behavior at high frequency as observed in Figure 3. In particular, in samples A and D the low frequency feature is more pronounced than in samples B and C. The proposed model is capable to capture all these characteristics and to satisfactorily fit all the four curves as reported in Figure 6. This indicates that the model is able not only to reproduce the frequency and $P_{O2}$ dependence as discussed in Section 3.2, but also to take into account different microstructural characteristics of the electrode. Table 2 reports the corresponding boundary layer length values $l_\delta$ as evaluated from the fitting of the experimental data.

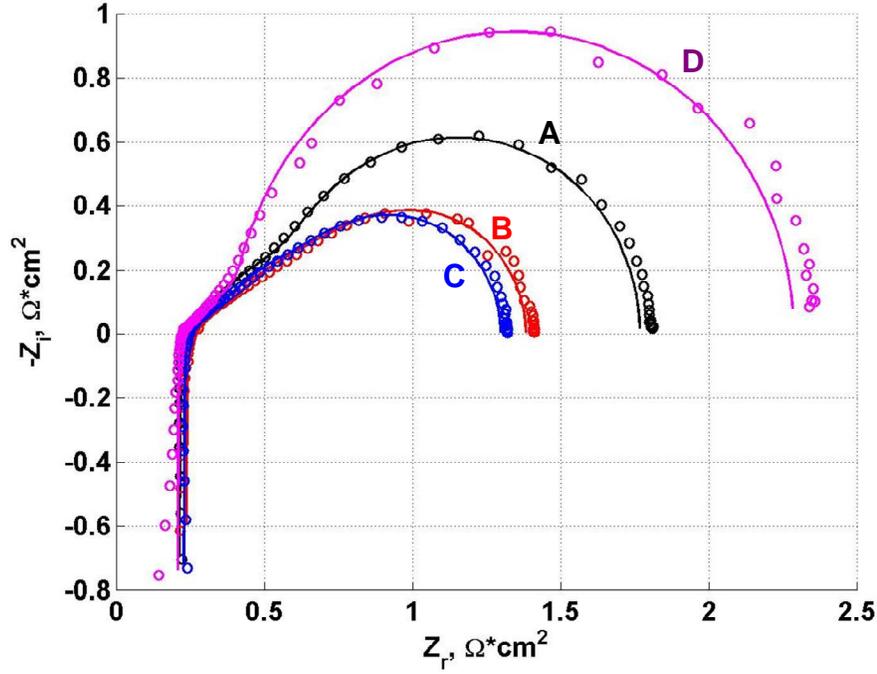

**Figure 6 Impedance spectra at OCV, 800ºC and 21% $P_{O2}$ for symmetric cells with different microstructures of the cathode functional layer. Experimental data are reported with marks, model simulations with solid lines.**

It is interesting to link the impedance behavior of the samples, especially their polarization resistance, to their different microstructural characteristics. According to the operating conditions adopted, that is, 800°C and no bias overpotential, the surface path of the ORR should be the dominant reaction mechanism [3,10]. In such a case, as reported in the Introduction Section, the current is expected to scale with the TPB length [11–13], that is, the cathode polarization resistance should scale inversely with $l_{TPB}$. Note that such a dependence is foreseen by the analytical model too, see Eqs. (11) and (12).

The TPB density $l_{TPB}$ of the four samples can be roughly estimated from the microstructural characteristics reported in Table 2. In particular, the TPB length per unit of electrode area is proportional to the number density of TPB points per unit area $n_{TPB}$ times the thickness of the cathode functional layer $t_{CFL}$:

$$l_{TPB} \propto n_{TPB} \cdot t_{CFL} \qquad (15)$$

However, the comparison between experimental data in Figure 6 and microstructural properties in Table 2 indicates that there is not a clear relationship between TPB density and polarization resistance. For example, cells B, C and D show almost the same $n_{TPB}$, however the polarization resistance of sample D is about twice larger according to Figure 6. Moreover, sample A has the lowest TPB density but its polarization resistance lies between samples B and D.

Three possible reasons can be formulated to explain this result:

- bi-dimensional SEM images may not allow a fair estimation of the volumetric TPB density of the samples, thus the values of $n_{TPB}$ reported in Table 2 may not be representative of the real electrode microstructure;

- since in SEM images it is not possible to distinguish between percolating and non-percolating triple phase boundaries, different percolation fractions may reduce the density of percolating TPB, which may significantly differ from what reported in Table 2;

- the TPB density may not be the main microstructural parameter that determines the cell performance.

Regarding the last point, the derivation of the model reported in Section 2 highlights that the polarization resistance is affected by other microstructural parameters, in particular by the surface diffusion length $L_s$ as indicated in Eqs. (11) and (12). The impact of the microstructural parameter $L_s$ on the interpretation of experimental results is discussed in the next Section.

## 4) Discussion: surface diffusion length and Thiele modulus

In the previous Section, the analysis of impedance spectra for different electrode microstructures revealed that the model is able to reproduce the experimental behavior but the cathode performance does not scale directly with the TPB density. In this Section, the role of another important microstructural parameter, that is, the surface diffusion length $L_s$, is discussed for a different interpretation of the same experimental data.

Table 2 reports the values of $L_s$ of the four samples as estimated from the SEM image analysis. Sample D, which gives the poorest performance in Figure 6, shows the lowest $L_s$. On the other hand, samples B and C have a larger $L_s$ and show smaller resistances. Finally, sample A, which shows an intermediate polarization resistance, has a midway $L_s$. This observation suggests that there is a strong correlation between the cathode polarization resistance and the surface diffusion length.

This analysis can be further refined by taking into account also the effect of the boundary layer length $l_\delta$, whose values, obtained through the fitting of Figure 6, are reported in Table 2. Sample D, which shows the highest polarization resistance, not only has the smallest $L_s$, it also has the largest $l_\delta$. Sample B has a larger $L_s$ than sample C, however the two samples show similar performance: this can be explained by noting that $l_\delta$ is larger in sample B than in sample C. Finally, while samples A and C have a comparable $L_s$, sample C performs better due to the smaller boundary layer length $l_\delta$.

Hence, this analysis suggests that the relative ranking of cathode resistances can be interpreted as a function of the ratio between $L_s$ and $l_\delta$, that is, as a function of the Thiele modulus $\phi$ defined in Eq. (14). Indeed, fixed the other parameters, Eq. (12) predicts an inverse relationship between the cathode polarization resistance $R_G$ and the Thiele modulus $\phi$, as reported in Figure 7. When the Thiele modulus is smaller than one, the resistance is relatively big and increases sharply as $\phi$ decreases. On the other hand, for $\phi > 1$ the resistance decreases more smoothly and gradually reaches a plateau. Notably, Figure 7 shows that such a dependence is quantitatively exhibited by the experimental data as well, which corroborates the validity of the model and of the interpretation summarized above.

This result confirms that the electrochemical performance of the four cathodes analyzed in this study is essentially ruled by the electrocatalytic activity of the LSM, that is, by the Thiele modulus. This indicates that, at least in the conditions investigated in this study, the surface diffusion length is the principal microstructural parameter to take into account rather than the TPB density. Thus, in order to reduce the cathode resistance, microstructural modifications should be undertaken to increase the surface diffusion length $L_s$. This is valid provided that the TPB density $l_{TPB}$ is not significantly decreased because, otherwise, an increase in polarization resistance is expected according to Eq. (12).

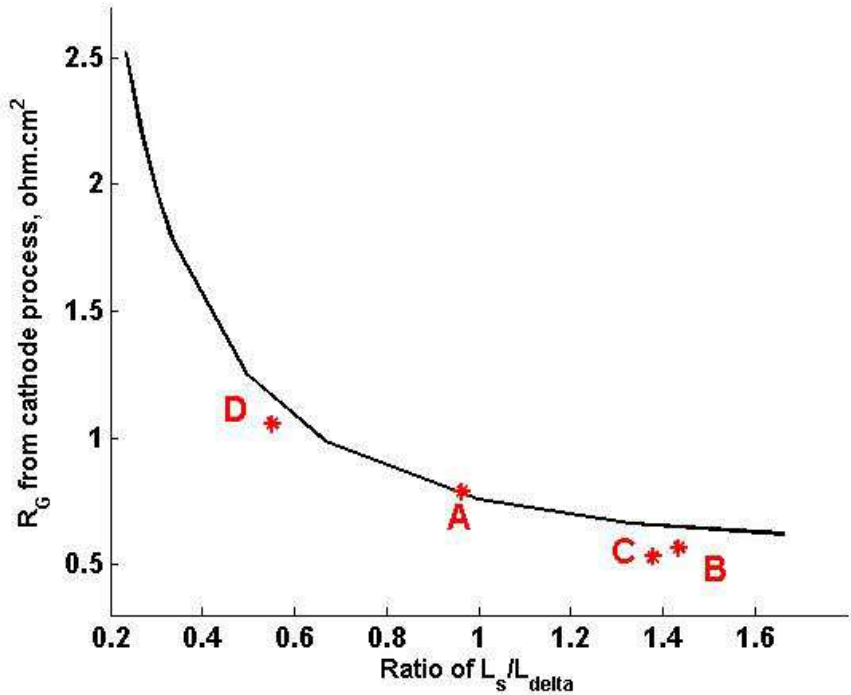

**Figure 7 Cathode polarization resistance $R_G$ as a function of the ratio between the surface diffusion length $L_s$ and the boundary layer length $l_\delta$ (i.e., the Thiele modulus) as predicted by the analytical model. The simulated operating conditions are 800 ºC and 21% of $P_{O2}$; all the model parameters are kept fixed while $L_s$ is varied. The experimental polarization resistances of the four samples are reported with marks.**

The important role of the Thiele modulus on the electrocatalytic activity of the electrode is schematically depicted in Figure 8. When the Thiele modulus is much larger than 1 (Figure 8a), the catalyst surface is well covered by the adsorbed species, thus the overall process is limited by the surface diffusion inside a thin boundary layer $l_\delta$. In this scenario, the catalyst shows good activity. On the other hand, when the Thiele modulus is much smaller than 1 (Figure 8b), the gas adsorption flux is too low to sustain the diffusive flux toward the reaction sites, thus the adsorption becomes the rate-determining step of the whole process. In such as case, since the adsorbed species quickly diffuse to the reaction site, the catalyst surface is barely covered by the adsorbed species throughout the surface boundary length. In this case, the catalyst shows a relatively bad activity.

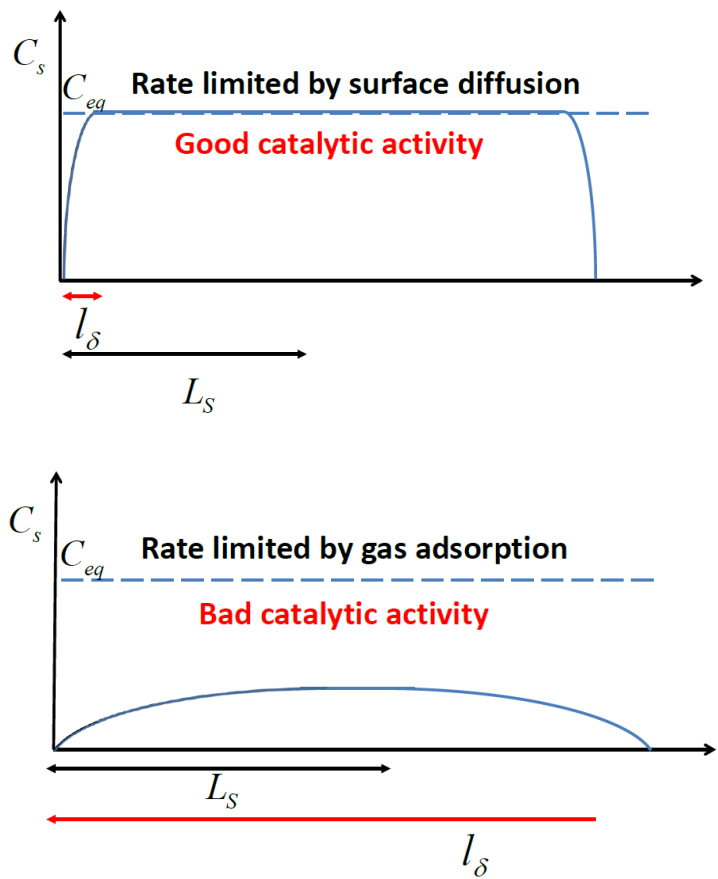

**Figure 8 Schematic illustration of the role of the Thiele modulus in electrocatalysis. (a) Thiele modulus much larger than 1 (b) Thiele modulus much smaller than 1.**

The dependence of the surface coverage on the Thiele modulus has a relevant impact on the impedance response of the electrode. Figure 9 shows the simulated impedance spectra of the cathode diffusion process $Z_G$ (Eq. (11)) for different Thiele moduli. When the Thiele modulus is small, that is, when the surface diffusion length $L_s$ is shorter than the boundary layer length $l_\delta$ as in Figure 8b, the impedance response of the cathode surface process produces a semicircle in the low frequency region of the Nyquist plot, followed by a small straight portion, with a 45° slope, in the high frequency range. The semicircular portion represents the chemical capacitor behavior associated to the species adsorbed on the whole catalyst surface, which acts as a buffer. This is the distinct feature of the slow adsorption process, which limits the whole surface process. As mentioned above, in such a case LSM shows a bad catalytic activity, indeed the polarization resistance, which is the intercept with the real axis at $\omega = 0$, is relatively high. When $L_s$ increases (i.e., when $\phi$ increases), the semicircle at low frequency becomes smaller and the curve assumes a depressed shape, approaching the infinite-length Gerischer limit [43]. This condition corresponds to the situation depicted in Figure 8a, in which molecular oxygen adsorbs onto a large catalyst surface and oxygen adatoms diffuse within a short boundary layer length toward the TPB. In such a case, the surface process is limited by surface diffusion and the catalyst shows good activity. Note that these two representative regimes, corresponding to small and large Thiele modulus, are experimentally reproduced by the four cathodes with different microstructures as reported in Figure 6.

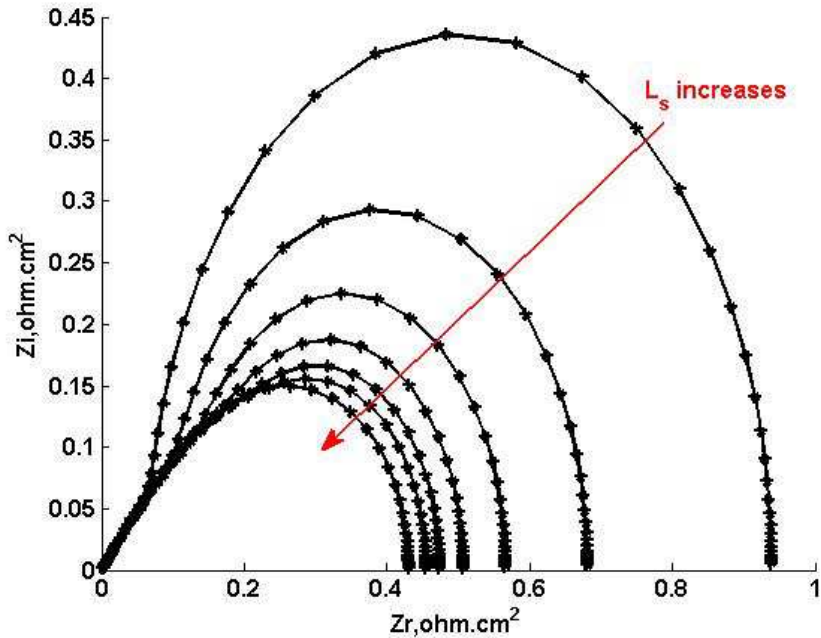

**Figure 9 Theoretical impedance response of the cathode surface diffusion process when the surface diffusion length $L_s$ is gradually changed. In this simulation, the boundary layer length is set to 0.20 μm.**

Finally, the electrocatalytic activity of LSM is investigated with respect to the adsorption equilibrium by using the model. In Figure 10 the resistance of the surface process $R_G$ is plotted as a function of the adsorption equilibrium constant $K$ according to Eq. (12). Figure 11 shows a similar plot by reporting the dependence of $R_G$ on the oxygen partial pressure $P_{O2}$. Note that, as a model simplification, in Figure 11 it is assumed that the surface properties of LSM (e.g., $D_s$, $k_{ads}$, $C_{max}$, etc.) do vary with $P_{O2}$, although in reality the perovskite stoichiometry of LSM is a function of the oxygen partial pressure [3,32].

In both Figures 10 and 11, the logarithmic plot of $1/R_G$ shows a volcano shaped curve with a maximum, which corresponds to the best cathode performance. Such a volcano shape is the graphical representation of the Sabatier principle of heterogeneous catalysis [50], which has been used in the molecular engineering of electrocatalysts for the ORR and other Faradaic reactions [51–53]. The principle states that the catalyst shows the maximum activity when its interaction with the reactant is neither too strong nor too weak. For the adsorption process simulated in this study, a weak interaction between oxygen and catalyst corresponds to a small value of $K$ or a low $P_{O2}$. In such a case, the equilibrium surface fraction $\theta_{eq}$ approaches 0 according to Eq. (6): since just a few oxygen adatoms cover the LSM surface, the whole adsorption-diffusion process proceeds slowly, resulting in a large resistance $R_G$. On the other hand, a strong interaction between oxygen and catalyst corresponds to a large value of $K$ or a high $P_{O2}$. In this situation most of the surface sites are occupied by oxygen adatoms, thus the reduced concentration of free sites slows down the adsorption kinetics as reported in Eq. (4), resulting in a large surface resistance $R_G$. Thus, the Sabatier principle predicts an optimum condition at intermediate values of $K$ or $P_{O2}$, for which the catalyst shows the maximum activity, resulting in the minimum resistance $R_G$. As shown in Figures 10 and 11, the model predictions satisfy the Sabatier principle of catalysis, thus further corroborating the analysis of the oxygen reduction reaction discussed in Sections 3 and 4.

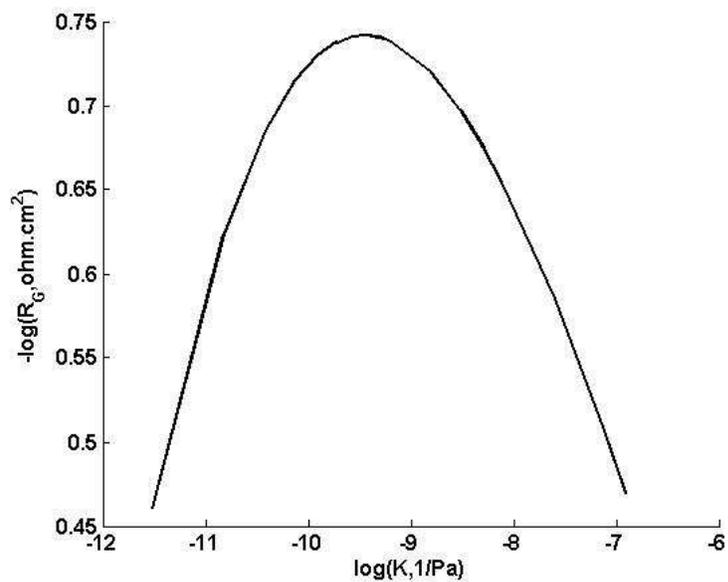

**Figure 10 The theoretical effect of the adsorption equilibrium constant *K* on the inverse of the resistance of the cathode surface adsorption-diffusion process *1/R$_G$* as calculated from Eq. (12). The scale is logarithmic for both the axes.**

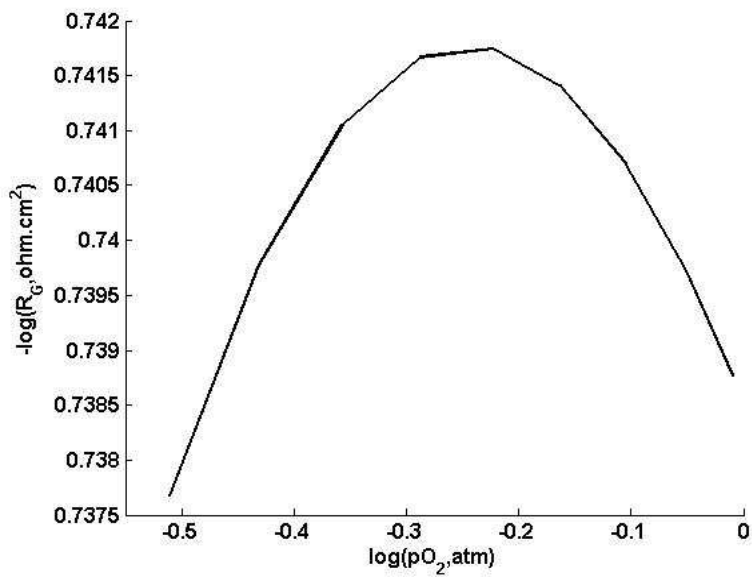

**Figure 11 The theoretical effect of the oxygen partial pressure *P$_{O2}$* on the inverse of the resistance of the cathode surface adsorption-diffusion process *1/R$_G$* as calculated from Eq. (12). The scale is logarithmic for both the axes.**

## 5) Conclusions

In this paper, a general model for the heterogeneous electrocatalysis of porous composite solid oxide fuel cell cathodes was developed. The model is based on dissociative Langmuir adsorption coupled with surface diffusion of adsorbed intermediates, in conjunction with fast charge transfer reactions. An analytical expression of the cathode impedance, which has the form of a finite-length Gerischer-type element, was derived. In contrast to the usual phenomenological description of the experimental spectra, our physics-based approach connects the functional form and parameters of the equivalent circuit with microscopic physical processes and material properties of the electrode, thus allowing for a thorough understanding of the system and its precise validation and testing.

The model was validated with experimental impedance data obtained in double-cathode symmetric cell configuration, showing its capability to reproduce the frequency and oxygen partial pressure dependences of composite LSM/YSZ cathodes. In addition, the analysis of cathodes with different microstructures allowed the quantitative correlation of the impedance response to the microstructural characteristics of the electrodes. The model revealed that the cathode resistance is controlled by an electrocatalytic dimensionless number, corresponding to the ratio between the surface diffusion length, which is a microstructural parameter corresponding to the semi-distance between two TPBs on the LSM surface, and the boundary layer length, which is a characteristic number of the adsorption-diffusion process. Model simulations indicated that the electrode performance can be improved by maximizing the surface diffusion length, provided that a sufficient TPB density is maintained. Finally, the model satisfied the Sabatier principle of catalysis by producing the typical volcano plots of catalytic activity as a function of the strength of the adsorption equilibrium. These results give general insights into the basic physics of heterogeneous electrocatalysis and provide practical guidance for the material and microstructural optimization of porous electrode interfaces for SOFC and other battery or fuel cell applications.

# Appendix: Linearization of the Nernst equation of reaction (3)

The application of the Nernst equation to the charge transfer reaction (3) yields the following relationship between potential difference and species activities [54]:

$$V = V^\Theta + \frac{RT}{2F} \ln\left(\frac{a_{O(s)} a_e^2}{a_{O(YSZ)}}\right) \quad \text{(A1)}$$

where $V$ represents the potential difference between LSM and YSZ, $a_{O(s)}$ the activity of oxygen adatoms, $a_e$ the electron activity, $a_{O(YSZ)}$ the activity of oxygen ions in the YSZ bulk and $\Theta$ denotes standard reference conditions. The activities of electrons and oxygen ions can be assumed to be constant during the charge transfer process [55], thus can be absorbed into $V^\Theta$. On the other hand, when an ideal solution of adatoms and free sites on the LSM surface is assumed, the activity of adsorbed oxygen adatoms results equal to $a_{O(s)} = \theta/(1-\theta)$ [54]. Thus, Eq. (A1) becomes:

$$V = V^\Theta + \frac{RT}{2F} \ln\left(\frac{\theta}{1-\theta}\right) \quad \text{(A2)}$$

When an external perturbation is imposed to the system, a surface coverage perturbation $\Delta\theta$ and a voltage perturbation $\Delta V$ are generated. The surface coverage perturbation $\Delta\theta$ is expressed as in Eq. (5), while the voltage perturbation $\Delta V$ is similarly defined as follows:

$$V = V_{eq} + \Delta V, \text{ with } \Delta V = \delta V \cdot e^{i(\omega t + \vartheta)} \text{ and } \delta V \ll V_{eq} \quad \text{(A3)}$$

where $V_{eq}$ represents the potential difference in equilibrium conditions, calculated according to Eq. (A2) as:

$$V_{eq} = V^\Theta + \frac{RT}{2F} \ln\left(\frac{\theta_{eq}}{1-\theta_{eq}}\right) \quad \text{(A4)}$$

The substitution of Eqs. (5) and (A3) into Eq. (A2) yields:

$$V_{eq} + \Delta V = V^\Theta + \frac{RT}{2F} \ln\left(\frac{\theta_{eq} + \Delta\theta}{1-\theta_{eq}-\Delta\theta}\right) \quad \text{(A5)}$$

which is reduced to Eq. (A6) upon subtraction of Eq. (A4):

$$\Delta V = \frac{RT}{2F} \ln\left( \frac{\theta^{eq} + \Delta\theta}{\theta^{eq}} \frac{1 - \theta^{eq}}{1 - \theta^{eq} - \Delta\theta} \right) \qquad \textbf{(A6)}$$

Eq. (A6) can be rearranged as follows:

$$\Delta V = \frac{RT}{2F} \ln\left( 1 + \frac{\Delta\theta}{\theta^{eq}} \right) - \frac{RT}{2F} \ln\left( 1 - \frac{\Delta\theta}{1 - \theta^{eq}} \right) \qquad \textbf{(A7)}$$

Finally, since $\Delta\theta/\theta_{eq} \ll 1$ and $\Delta\theta/(1-\theta_{eq}) \ll 1$, the natural logarithms are expanded in Taylor series to the first order of $\Delta\theta$ to yield:

$$\Delta V = \frac{RT}{2F} \frac{\Delta\theta}{\theta^{eq}} + \frac{RT}{2F} \frac{\Delta\theta}{1 - \theta^{eq}} = \frac{RT}{2F} \Delta\theta \left( \frac{1}{\theta^{eq}} + \frac{1}{1 - \theta^{eq}} \right) \Rightarrow \Delta\theta = \theta_{eq}(1 - \theta_{eq}) \frac{2F}{RT} \Delta V \qquad \textbf{(A8)}$$

which is exactly Eq. (8a) as reported in Section 2.